\documentclass[12pt]{iopart}
\usepackage{epsfig}

\newcommand{\agt}{\,\rlap{\lower3.7pt\hbox{$\mathchar\sim$}}
\raise1pt\hbox{$>$}\,}
\newcommand{\alt}{\,\rlap{\lower3.7pt\hbox{$\mathchar\sim$}}
\raise1pt\hbox{$<$}\,}

\long\def\dump#1{}

\begin{document}

\title[Axions]{Axions---Motivation, limits and
searches\footnote{Contribution to Proceedings of {\it IRGAC 2006:
2nd International Conference on Quantum Theories and Renormalization
Group in Gravity and Cosmology\/} (11--15 July 2006, Barcelona,
Spain)}}

\author{G G Raffelt}

\address{Max-Planck-Institut f\"ur Physik
 (Werner-Heisenberg-Institut),
 F\"ohringer Ring 6, 80805 M\"unchen,
 Germany}
\ead{raffelt@mppmu.mpg.de}

\begin{abstract}
The axion solution of the strong CP problem provides a number of
possible windows to physics beyond the standard model, notably in the
form of searches for solar axions and for galactic axion dark matter,
but in a broader context also inspires searches for axion-like
particles in pure laboratory experiments. We briefly review the
motivation for axions, astrophysical limits, their possible
cosmological role, and current searches for axions and axion-like
particles.
\end{abstract}

\section{Introduction}                        \label{sec:introduction}

Quantum chromodynamics is a CP-violating theory, implying that the
neutron should have a large electric dipole moment, far in excess of
experimental limits~\cite{Baker:2006ts}. One elegant solution of
this ``strong CP problem'' was proposed by Peccei and Quinn (PQ) who
postulated a new global U(1) symmetry that is spontaneously broken
at some large energy scale, allowing for the dynamical restoration
of the CP symmetry~\cite{Peccei:1977hh, Peccei:1977ur}.
Weinberg~\cite{Weinberg:1977ma} and Wilczek~\cite{Wilczek:1977pj}
realized that an inevitable consequence of this mechanism is a new
pseudoscalar boson, the axion, which is the Nambu-Goldstone boson of
the PQ symmetry. The axion solution would open a new low-energy
window to high-energy physics.  String theory provides another
motivation in that axion-like fields typically appear so that
plausibly one of them could play the role of the QCD
axion~\cite{Svrcek:2006yi}.

The PQ symmetry is explicitly broken at low energies by instanton
effects so that the axion acquires a small mass,
\begin{equation}\label{eq:axmass}
m_a=\frac{z^{1/2}}{1+z}\,\frac{f_\pi m_\pi}{f_a} =\frac{6.0~{\rm
eV}}{f_a/10^6~{\rm GeV}}\,,
\end{equation}
where $f_a$ is the axion decay constant or PQ scale that governs all
axion properties, $f_\pi=93$~MeV is the pion decay constant, and
$z=m_u/m_d$ is the mass ratio of up and down quarks. We will follow
the previous axion literature and assume the canonical value $z=0.56$
\cite{Gasser:1982ap, Leutwyler:1996qg}, although it could vary in the
range 0.3--0.6 \cite{Yao:2006px}.

Axions were not found in early searches, ruling out ``standard
axions'' where $f_a$ would have been related to the electroweak
scale $f_{\rm EW}$. ``Invisible'' axion models with $f_a\gg f_{\rm
EW}$ were quickly constructed~\cite{Kim:1979if, Shifman:1979if,
Zhitnitsky:1980tq, Dine:1981rt, Kim:1986ax} and quickly recognized
to be far from inconsequential because they can provide the dark
matter of the universe~\cite{Preskill:1982cy, Abbott:1982af,
Dine:1982ah, Davis:1986xc} and can be searched in realistic
experiments~\cite{Sikivie:1983ip, Bradley:2003kg, Asztalos:2003px,
Duffy:2006aa}.

The properties of axions are closely related to those of neutral
pions. One generic property is a two-photon interaction that plays a
key role for most searches,
\begin{equation}
{\cal L}_{a\gamma}= \frac{1}{4}g_{a\gamma} F_{\mu\nu}\tilde
F^{\mu\nu}a =-g_{a\gamma}\,{\bf E}\cdot{\bf B}\,a\,.
\end{equation}
Here, $F$ is the electromagnetic field-strength tensor, $\tilde F$
its dual, and ${\bf E}$ and ${\bf B}$ the electric and magnetic
fields, respectively. The coupling constant is
\begin{equation}\label{eq:axionphoton}
 g_{a\gamma}=\frac{\alpha}{2\pi f_a}
 \left(\frac{E}{N}-\frac{2}{3}\,\frac{4+z}{1+z}\right)
 =\frac{\alpha}{2\pi}
 \left(\frac{E}{N}-\frac{2}{3}\,\frac{4+z}{1+z}\right)
 \frac{1+z}{\sqrt{z}}\,\frac{m_a}{m_\pi f_\pi}\,,
\end{equation}
where, $E$ and $N$, respectively, are the electromagnetic and color
anomaly of the axial current associated with the axion field. In
grand unified models, e.g.\ the DFSZ model~\cite{Zhitnitsky:1980tq,
Dine:1981rt}, axions couple to ordinary quarks and leptons, implying
$E/N=8/3$, whereas in the KSVZ model~\cite{Kim:1979if,
Shifman:1979if} $E/N=0$. While these cases are often used as generic
examples, in general $E/N$ is not known so that for fixed $f_a$ a
broad range of $g_{a\gamma}$ values is possible~\cite{Cheng:1995fd}.
Still, barring fine-tuned cancelations, $g_{a\gamma}$ scales from
the corresponding pion interaction by virtue of
Eq.~(\ref{eq:axionphoton}). Taking the model-dependent factors to be
of order unity defines the ``axion line'' in the $m_a$-$g_{a\gamma}$
plane.

Axions transform into photons and vice versa in external $B$
fields~\cite{Sikivie:1983ip}, a phenomenon similar to neutrino
oscillations~\cite{Raffelt:1987im}.  This effect can have intriguing
applications, although the relevant range of $g_{a\gamma}$ and $m_a$
typically is far away from the axion line because for a given
$g_{a\gamma}$ the implied~$m_a$ usually suppresses conversion effecs
by energy-momentum conservation. Therefore, beginning with
``arions'' \cite{Anselm:1981aw, Anselm:1982ip}, the axion relation
between $g_{a\gamma}$ and $m_a$ was often taken to be relaxed,
postulating nearly massless axion-like particles for which ``ALPs''
has recently become standard coinage. While
CAST~\cite{Andriamonje:2004hi} (solar axions) and
ADMX~\cite{Asztalos:2003px} (galactic dark matter) are the only
experiments that probe ALPs on the axion line, we will briefly
review other searches, notably the puzzling signal in the PVLAS
experiment~\cite{Zavattini:2005tm}.

Section~\ref{sec:axionphoton} is devoted to searches for axions and
axion-like particles based on their two-photon interaction and the
relevant astrophysical limits. In Sec.~\ref{sec:astrolimits} other
astrophysical limits are summarized.  Section~\ref{sec:cosmology} is
devoted to the possible cosmological role of axions before turning to
a summary in Sec.~\ref{sec:summary}.

\section{Axion-photon interaction: Searches, limits, applications}
\label{sec:axionphoton}

\subsection{Searches for solar axions}

Particles with a two-photon vertex, including neutral pions and
gravitons besides the hypothetical axions, can transform into
photons in external electric or magnetic fields, an effect first
discussed by Primakoff in the early days of pion
physics~\cite{Primakoff}. Therefore, stars produce these particles
from thermal photons in the fluctuating electromagnetic fields of
the stellar plasma~\cite{Dicus:fp}. Calculating the solar axion flux
is straightforward except for the proper inclusion of screening
effects~\cite{Raffelt:1985nk, Altherr:1993zd}. The transition rate
for a photon of energy $E$ into an axion of the same energy (recoil
effects are neglected) is~\cite{Raffelt:1987np}
\begin{equation}\label{eq:Gamma-ag}
\Gamma_{\gamma\to a}= \frac{g_{a\gamma}^2T\kappa_{\rm s}^2}{32 \pi}
\bigg[\bigg(1+\frac{\kappa_{\rm s}^2}{4E^2}\bigg)
\ln\bigg(1+\frac{4E^2}{\kappa_{\rm s}^2}\bigg)-1\bigg]\,,
\end{equation}
where $T$ is the temperature (natural units with $\hbar=c=k_B=1$ are
used). The screening scale in the Debye-H\"uckel approximation~is
\begin{equation}
\kappa_{\rm s}^2= \frac{4\pi\alpha}{T}\biggl(n_e+\sum_{\rm nuclei}
Z_j^2n_j\biggr),
\end{equation}
where $n_e$ is the electron density and $n_j$ that of the $j$-th ion
of charge $Z_j$. Near the solar center $\kappa_{\rm s}\approx 9$~keV
and $(\kappa_{\rm s}/T)^2\approx12$ is nearly constant throughout
the Sun. The axion flux at Earth, based on a standard solar model,
is well approximated by
\begin{equation}\label{eq:bestfit}
 \frac{d\Phi_a}{dE}=g_{10}^2\,\,6.0\times
 10^{10}~{\rm cm}^{-2}~{\rm s}^{-1}~{\rm keV}^{-1}\,
 E^{2.481}\, {\rm e}^{-E/1.205}\,,
\end{equation}
where $E$ is in keV and $g_{10}=g_{a\gamma}/(10^{-10}~{\rm
GeV}^{-1})$. The integrated flux parameters are
\begin{equation}
\Phi_a=g_{10}^2\, 3.75\times10^{11}~{\rm cm}^{-2}~{\rm s}^{-1}
\hbox{\quad and\quad}
L_a=g_{10}^2\,1.85\times 10^{-3} L_\odot\,.
\end{equation}
The maximum of the distribution is at $3.0$~keV, the average energy
is $4.2~{\rm keV}$.

This flux can be searched with the inverse process where an axion
converts into a photon in a macroscopic $B$ field, the ``axion
helioscope'' technique~\cite{Sikivie:1983ip}.  One would look at the
Sun through a ``magnetic telescope'' and place an x-ray detector at
the far end.  The conversion can be coherent over a large propagation
distance and is then pictured as a particle oscillation
effect~\cite{Raffelt:1987im}. The conversion probability is
\begin{equation}
P_{a\to\gamma}=\left(\frac{g_{a\gamma}B}{q}
\right)^2\sin^2\left(\frac{qL}{2}\right)\,,
\end{equation}
where $L$ is the path length and $q$ the axion-photon momentum
difference; in vacuum it is $q=m_a^2/2E$.  For $q L\alt 1$ the
oscillation length exceeds $L$. For $q L\ll 1$ we have
$P_{a\to\gamma}=(g_{a\gamma}B L/2)^2$, implying an x-ray flux of
\begin{equation}
\Phi_\gamma=0.51~{\rm cm^{-2}~day^{-1}}\,g_{10}^4\,
\left(\frac{L}{9.26~\rm m}\right)^2 \left(\frac{B}{9.0~\rm
T}\right)^2\,.
\end{equation}
For $qL\agt 1$ this rate is reduced by the momentum mismatch. A
low-$Z$ gas would provide a refractive photon mass $m_\gamma$ so
that $q=|m_\gamma^2-m_a^2|/2E$. For $m_a\approx m_\gamma$ the
maximum rate can thus be restored~\cite{vanBibber:1988ge}.

Early helioscope searches were performed in
Brookhaven~\cite{Lazarus:1992ry} and Tokyo~\cite{Moriyama:1998kd,
Inoue:2002qy}. Solar axions could also transform in electric crystal
fields, but the limits obtained by SOLAX~\cite{Avignone:1997th},
COSME~\cite{Morales:2001we}, and DAMA~\cite{Bernabei:ny} are less
restrictive and require an axion luminosity exceeding the
helioseismological constraint $g_{10}\alt10^{-9}~{\rm
GeV}^{-1}$~\cite{Schlattl:1998fz}.

\begin{figure}[b]
\begin{center}
\epsfig{file=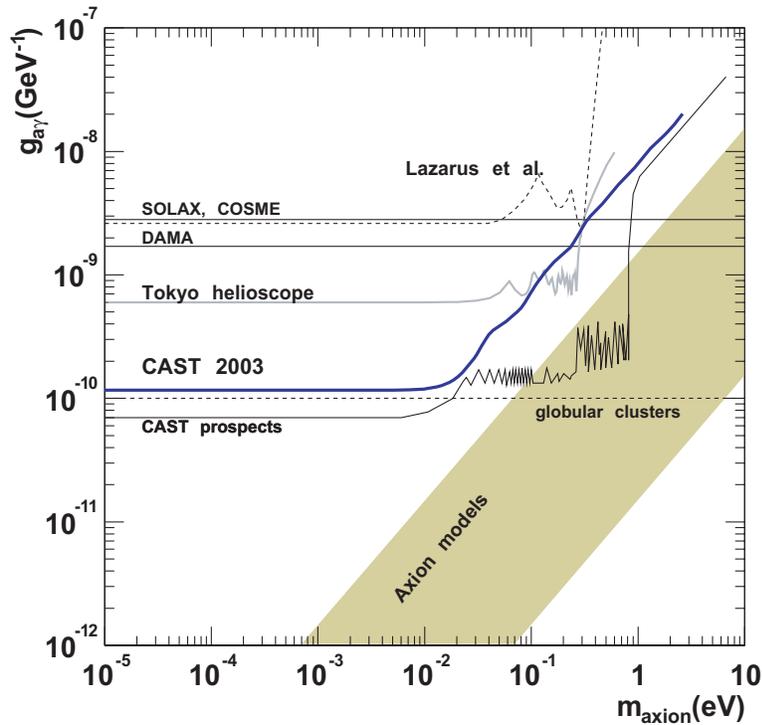,width=10cm}
\end{center}
\caption{Exclusion limit (95\% CL) from CAST 2003 compared with
other constraints discussed in the
text~\protect\cite{Andriamonje:2004hi}. The shaded band represents
typical theoretical models (``axion line''). Also shown is the
future sensitivity foreseen in the CAST proposal.} \label{fig:cast}
\end{figure}

The first helioscope that can actually reach the ``axion line'' is
the CERN Axion Solar Telescope (CAST), using a refurbished LHC test
magnet with $B=9.0~\rm T$ and two pipes of length $L=9.26~\rm m$.
The magnet is mounted with $\pm 8 ^\circ$ vertical movement,
allowing for observation of the Sun for 1.5 h at both sunrise and
sunset.  CAST operated for about 6~months in 2003. The
non-observation of a signal above background in all three detectors
leads to the exclusion range shown in Fig.~\ref{fig:cast}. In the
mass range $m_a\alt0.02$~eV the new limit
is~\cite{Andriamonje:2004hi}
\begin{equation}\label{eq:castlimit}
g_{a\gamma}<1.16\times 10^{-10}~{\rm GeV}^{-1}
\hbox{\quad(95\% CL)}\,.
\end{equation}
The data taken in 2004 do not show any apparent signal above
background, leading to limits comparable to the sensitivity
prospected in Fig.~\ref{fig:cast}.

In the ongoing Phase~II, a variable-pressure helium filling provides
an effective photon mass. The vapor pressure of He$^4$ at 1.8~K, the
magnet's operating temperature, allows one to reach $m_a$ up to
0.26~eV. In the next phase, He$^3$ will be used because its higher
vapor pressure allows one to reach 0.8~eV. Yet higher masses require
an isolating gas cell in the bore where He$^3$ at 5.4~K would allow
one to reach $m_a$ of~1.4~eV.

For $m_a\alt10^{-4}$, a competitive method to search for solar
axions could be to look with an x-ray satellite at the Sun when it
is shadowed by the Earth and search for x-rays from the axion-photon
conversion in the Earth's magnetic field~\cite{Davoudiasl:2005nh}. A
related idea is to look for high-energy $\gamma$-rays, e.g.\ with
the GLAST satellite, that ``pass through the Sun'' by magnetic
conversion and back conversion~\cite{Fairbairn:2006hv}.

\subsection{Do axions escape from the Sun?}

CAST can detect axions only if they actually escape from the Sun.
Their mean free path (mfp) against the Primakoff process is the
inverse of Eq.~(\ref{eq:Gamma-ag}). As an example we consider 4~keV
axions and note that at the solar center $T\approx 1.3$~keV and
$\kappa_{\rm s}\approx9$~keV. The axion mfp is then
$\lambda_a\approx g_{10}^{-2}\,6\times10^{24}~{\rm cm}\approx
g_{10}^{-2}\,8\times10^{13}\,R_{\odot}$, or about $10^{-3}$ of the
radius of the visible universe. Therefore, $g_{a\gamma}$ would have
to be more than $10^7$ larger than the CAST limit for axions to be
re-absorbed in the Sun. Even in this extreme case they are not
harmless because they would carry the bulk of the energy flux that
otherwise is carried by photons. The mfp of low-mass particles that
are trapped in the Sun should be shorter than that of photons (about
10~cm near the solar center) to avoid a dramatic modification of the
solar structure~\cite{Raffelt:1988rx}. This requirement is so
extreme that for anything similar to axions the possibility of
re-absorption is not a serious possibility.

\subsection{Globular cluster stars}

The most restrictive astrophysical limit on $g_{a\gamma}$ arises
from globular cluster stars. A globular cluster is a gravitationally
bound system of stars that formed at the same time and thus differ
primarily in their mass. A globular cluster provides a homogeneous
population of stars, allowing for detailed tests of
stellar-evolution theory. The stars surviving since formation have
masses somewhat below $1\,M_\odot$. In a color-magnitude diagram,
where one plots essentially the surface brightness vs.\ the surface
temperature, stars appear in characteristic loci, allowing one to
identify their state of evolution.

The stars on the horizontal branch (HB) have reached helium burning
where their core (about $0.5\,M_\odot$) generates energy by fusing
helium to carbon and oxygen with a core-averaged energy release of
about $80~\rm erg~g^{-1}~s^{-1}$.  On the other hand, the average
Primakoff energy loss is approximately $g_{10}^2\,30~\rm
erg~g^{-1}~s^{-1}$. The main effect would be an acceleration of the
consumption of helium and thus a reduction of the HB lifetime by a
factor $80/(80+30\,g_{10}^2)$, i.e., by about 30\% for $g_{10}=1$.
Number counts in 15~globular clusters~\cite{Buzzoni:1983} reveal
typically 100 HB stars in the used fields of view. This is compared
with the number of red giants (evolutionary phase before helium
ignition, but after core-exhaustion of hydrogen) and shows that the
HB lifetime in any one cluster is established within about 20--40\%.
Note that the Primakoff losses are much smaller for red giants so
that their lifetime is not reduced by Primakoff emission.

Compounding the results of all 15 clusters, the helium-burning
lifetime agrees with expectations to within about 10\%
\cite{Raffelt:1996wa, Raffelt:1999tx}. Of course, with modern data
these results likely could be improved. Either way, a reasonably
conservative limit is (Figs.~\ref{fig:cast} and~\ref{fig:exclusion})
\begin{equation}
g_{a\gamma}<10^{-10}~{\rm GeV}^{-1}\,.
\end{equation}
It is comparable to the CAST limit Eq.~(\ref{eq:castlimit}), but
applies for higher masses because the relevant temperature is about
10~keV so that significant threshold effects only begin at about
$m_a\agt 30$~keV. For QCD axions the coupling increases with mass so
that the limit reaches to even larger masses.

In the helium-burning core, convection and semi-convection dredges
helium fuel to the burning site near the core's center so that
25--30\% of all helium is burnt during the HB phase. Therefore,
while the standard theoretical predictions depend on a
phenomenological treatment of convection, there is limited room for
additional energy supply, even if the treatment of convection were
grossly incorrect.

\subsection{ALP-photon conversion in astrophysical magnetic fields}

Large-scale magnetic fields are ubiquitous in astrophysics so that
photon-axion conversions could be of interest. In practice, the
energy-momentum mismatch typically prevents this effect from being
important anywhere near the ``axion line.'' On the other hand,
``axion-like particles'' (ALPs) with smaller masses may well exist
and could then have intriguing consequences. Examples are the
polarization of radio galaxies~\cite{Harari:1992ea} and
quasars~\cite{Hutsemekers:2005iz, Gnedin:2006fq}, the diffuse x-ray
background~\cite{Krasnikov:1996bm}, or ultra-high energy cosmic rays
if they have a photon component ~\cite{Gorbunov:2001gc,
Csaki:2003ef}.

One interesting cosmological consequence is the dimming of distant
sources by photon-ALP conversion in intergalactic $B$
fields~\cite{Csaki:2001yk, Csaki:2001jk, Deffayet:2001pc,
Grossman:2002by, bakuI, Ostman:2004eh, Csaki:2004ha, Mirizzi:2005ng,
Song:2005af}, for a review see Ref.~\cite{Mirizzi:2006zy}. The
apparent dimming of distant SNe~Ia is usually attributed to
accelerated cosmic expansion. All things considered, the required
dimming cannot be caused by photon-ALP conversion, but this effect
may still figure for a detailed interpretation of the cosmic
equation of state implied by current and future SN~Ia data.

Preventing excessive suppression of photon-ALP conversion in this
context requires $m_a\alt 10^{-16}$--$10^{-15}$~eV, i.e., the ALP
mass should be smaller than the photon plasma frequency in the
intergalactic medium. For such small masses, the absence of
$\gamma$-rays from SN~1987A implies
$g_{a\gamma}\alt1\times10^{-11}~{\rm GeV}^{-1}$, valid for
$m_a\alt10^{-9}$~eV \cite{Brockway:1996yr, Grifols:1996id}. The ALPs
would have been emitted in SN~1987A by the Primakoff effect and
converted in the galactic magnetic field. A similar argument applied
to the nearby red giant Betelgeuse yields a slightly less
restrictive limit~\cite{Carlson:1995wa}, but of course depends on
fewer assumptions. ALP-photon conversion in stellar magnetic fields
is also possible, notably in the magnetic fields of Sun
spots~\cite{Carlson:1995xf} or the strong magnetic fields of
pulsars~\cite{Dupays:2005xs, Lai:2006af}, although at present no new
limit or positive signature is available from these considerations.

\subsection{Dichroism and birefringence in laboratory
magnetic fields}

Magnetic fields induce vacuum birefringence in that electromagnetic
radiation with polarization parallel ($\parallel$) or perpendicular
($\perp$) to $B$ has a different refractive index induced by a
one-loop QED correction to Maxwell's
equations~\cite{Heisenberg:1935qt, Adler:1971wn}. This vacuum
Cotton-Mouton effect has never been measured, but can be searched in
the laboratory by propagating a linearly polarized laser beam in a
strong $B$ field and measure a small ellipticity that
develops~\cite{Iacopini:1979ci}.

ALPs, axions, neutral pions, and other particles with a two-photon
vertex also contribute to this effect in that the photon mixes with
them in a transverse $B$ field~\cite{Maiani:1986md, Raffelt:1987im}.
If sufficiently light, these particles also cause vacuum dichroism,
i.e., the $\parallel$ and $\perp$ polarization states are absorbed
differently due to photon-ALP conversion. The observable effect is a
rotation of the plane of polarization. A huge effect of this sort
has been reported by the PVLAS
collaboration~\cite{Zavattini:2005tm}, where the amplitude of the
observed dichroism is roughly $10^4$ times larger than the expected
QED ellipticity amplitude. An instrumental origin has not been
identified, but further tests and modifications of the experiment
will be performed.

In PVLAS, the superconducting magnet is rotated to modulate the
signal because a static dichroism or birefringence is overwhelmed by
residual instrumental effects. It has been proposed that the $B$
field rotation mixes sidebands to the primary laser frequency with a
separation corresponding to the rotation
frequency~\cite{Mendonca:2006pg}. While such effects would in
principle occur, the quantitative treatment in this paper vastly
overestimates the effect; not even the use of electromagnetic units
is correct. The slow magnet rotation does not cause a $10^4$ fold
enhancement of the QED effect.

Still, the exciting possibility that an ALP has been detected is
damped by the extreme requirements on its
properties~\cite{Zavattini:2005tm, Cameron:1993mr}
\begin{equation}
 g_{a\gamma}=\hbox{2--5}\times10^{-6}~{\rm GeV}^{-1}
 \hbox{\qquad and\qquad}
 m_a=\hbox{1--1.5}~{\rm meV}\,.
\end{equation}
With this coupling strength, the Sun's ALP luminosity would exceed
its photon emission by more than a factor of $10^6$ and thus could
live only for 1000~years. A number of models have been proposed to
circumvent this problem in that the ALP properties and interactions
could strongly depend on the environment in the Sun or on
energy~\cite{Masso:2005ym, Masso:2006gc, Antoniadis:2006wp,
Mohapatra:2006pv, Jaeckel:2006xm, Redondo:2006xx}. It has become
clear that to square the particle interpretation of PVLAS with the
astrophysical limits is a tall order. Therefore, barring the
identification of an instrumental effect that causes the PVLAS
signature, the best hope is to test this hypothesis directly,
notably in a new generation of ``shining light through walls''
experiments~\cite{Rabadan:2005dm, Kotz:2006bw, Lindner:2006,
Cantatore:2006, Afanasev:2006cv, Afanasev:2006, Battesti:2006,
Pugnat:2006}.

\section{Other Astrophysical Limits}
\label{sec:astrolimits}

\subsection{Axion-Electron Interaction}

Besides their generic interaction with photons, axions also interact
with fermions $\Psi$ by a derivative interaction of the form
\begin{equation}
 {\cal L}=\frac{C}{2f_a}\,
 \bar\Psi\gamma_5\gamma_\mu\Psi\,\partial^\mu a
 \hbox{\qquad or\qquad}
 {\cal L}=-{\rm i}\,\frac{C m}{f_a}\,
 \bar\Psi\gamma_5\Psi\,a\,,
\end{equation}
where $C$ is a model-dependent numerical factor and $m$ the fermion
mass. The pseudoscalar form is equivalent when only a single axion
is attached to the fermion line in a given Feynman graph. Subtleties
for axion-nucleon interactions in a nuclear medium arise because of
the axion-pion mixing~\cite{Carena:1988kr}. The pseudoscalar form
suggests to quantify the interaction in terms of a dimensionless
Yukawa coupling $g=C m/f_a$ and associated ``fine structure
constant'' $\alpha_a=g^2/4\pi$.

In stellar plasmas, axions can be emitted by the Compton process
$\gamma+e\to e+a$, bremsstrahlung $e+Ze\to Ze+e+a$, bound-free and
free-bound transitions, or pair annihilation $e^++e^-\to\gamma+a$.
For the Sun, $L_a=\alpha_a\,6.0\times10^{21}\,L_\odot$ with 25\% due
to the Compton process~\cite{Raffelt:1985nk}. Helioseismology
conservatively implies that a new energy-loss channel cannot exceed
20\% of $L_\odot$ \cite{Schlattl:1998fz}, implying
$\alpha_a<3\times10^{-23}$. Avoiding excessive white-dwarf
cooling~\cite{Raffelt:1985nj} or a delay of helium ignition in
globular-cluster stars~\cite{Raffelt:1996wa, Raffelt:1999tx} implies
$\alpha_a\alt1\times10^{-26}$. In the DFSZ model we have
$C_e=\cos^2\beta/3$, assuming three families of fermions. For
$\cos^2\beta=1$ the exclusion range is shown in
Fig.~\ref{fig:exclusion}. Near the limit, axion emission can affect
the cooling speed and thus the period decrease of ZZ Ceti stars
(pulsating white dwarfs), although a previously measured anomalous
period decrease of the star G117-B15A has disappeared in the light
of more recent analyses~\cite{Isern:2003xj}.

\subsection{Axion-Nucleon Interaction}

The axion-electron interaction provides extremely restrictive
limits, but in hadronic axion models this coupling does not exist at
tree level. Axions are a QCD phenomenon, so their interactions with
pions, nucleons and photons are far more generic than those with
electrons. Unfortunately, the astrophysical arguments leading to the
electron coupling cannot be trivially recycled because axions couple
to the nucleon spin and thus at low energies do not interact with
$\alpha$ particles. Therefore, processes of the sort
$\gamma+\alpha\to\alpha+a$ do not occur, although they provide
limits on the scalar or vector coupling of new
bosons~\cite{Grifols:1986fc, Grifols:1988fv}.

However, the Sun largely consists of hydrogen so that axions are
produced by the Compton-like process $\gamma+p\to p+a$. The cross
section is $\sigma=(4\pi/3)\,\alpha\alpha_a E^2/m^4$ with $m$ the
fermion mass, i.e., the energy loss rate caused by protons relative
to electrons is suppressed by a factor
$(m_e/m_p)^4=0.9\times10^{-13}$. Taking the proton density in the
Sun to be 1/2 that of electrons, using
$L_a=\alpha_a\,1.5\times10^{21}\,L_\odot$ for the solar axion
luminosity from the electron Compton process, and observing that
helioseismology allows at most a 20\% exotic energy loss, we find
$\alpha_a\alt 3\times10^{-9}$ for the axion-proton interaction.

\begin{figure}[b]
\begin{center}
\epsfig{file=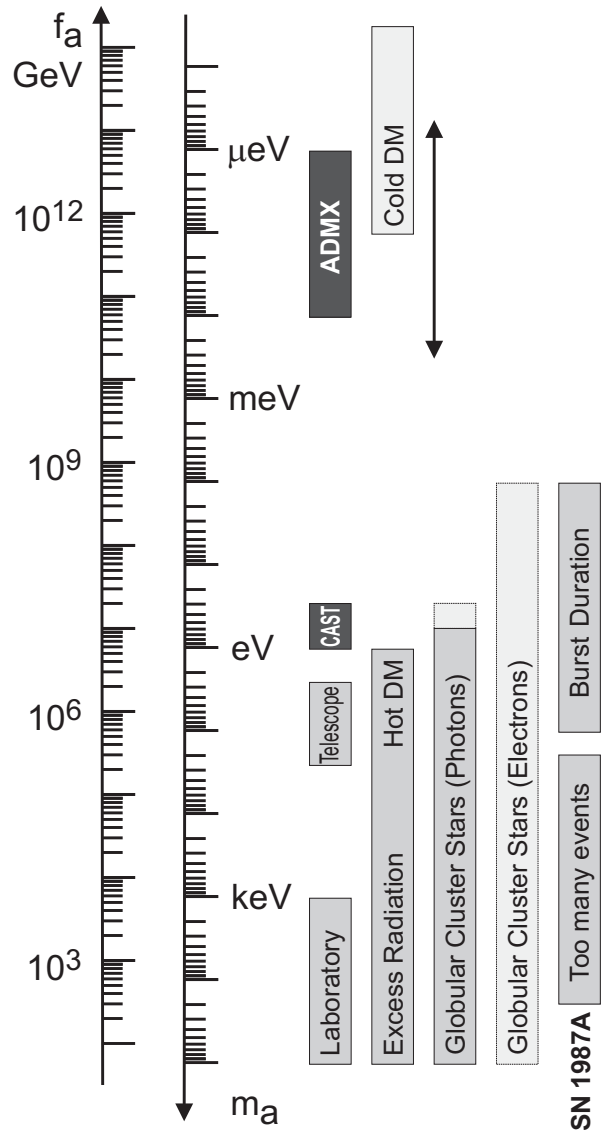,height=15cm}
\end{center}
\caption{Axion limits and foreseen search ranges as discussed in the
text. A light-grey ``exclusion bar'' means that it strongly depends
on cosmological or axion-model assumptions. The black bars indicate
ongoing or near-future axion searches.}\label{fig:exclusion}
\end{figure}

Much more restrictive limits arise from the observed neutrino burst
of supernova (SN) 1987A~\cite{Ellis:1987pk, Raffelt:1987yt,
Turner:1987by, Mayle:1987as, Mayle:1989yx, Burrows:1988ah,
Burrows:1990pk, Janka:1995ir, Keil:1996ju, Hanhart:2000ae}. The
burst lasted for about 10~s, in agreement with the usual picture of
neutrino trapping and slow diffusive energy loss of the collapsed SN
core. Excessive axion losses would have shortened the burst
duration. On the other hand, if axions interact so strongly that
their mfp is similar to that of neutrinos, they will not much affect
the neutrino signal. The results from different numerical
simulations and treatments agree reasonably
well~\cite{Raffelt:1996wa, Raffelt:1999tx} and lead to the
``consolidated'' exclusion range shown in Fig.~\ref{fig:exclusion}.

Early papers found more restrictive limits, but they overestimated
the emission rate. The main process is nucleon-nucleon
bremsstrahlung. Axions couple to the nucleon spin so that only
spin-nonconserving interactions contribute, i.e., the nuclear tensor
force. If it is modeled by one-pion exchange, the interaction rate
is so large that axions emitted from different collisions would
destructively interfere, leading to a suppression of the
Landau-Pomeranchuk-Migdal type~\cite{Janka:1995ir}. Later it was
noted that one-pion exchange may be a poor model for the relevant
conditions and the bremsstrahlung rate was calculated on the basis
of measured nuclear phase shifts, leading to a much smaller emission
rate~\cite{Hanhart:2000ae}. However, this result applies only in the
low-energy limit. Either way, the ``naive'' bremsstrahlung rate used
in early studies is reduced. Since the SN~1987A limits suffer from
these and other nuclear-physics uncertainties, but also from the
sparse experimental data, it is not easily possible to quantify an
objective error estimate.

If axions interact so strongly that they are trapped more
efficiently than neutrinos, still a significant flux will be
emitted. It interacts with oxygen in the water Cherenkov detectors
that have recorded the SN~1987A signal and could have mimicked some
of the signal~\cite{Engel:1990zd}. This argument excludes another
range of axion parameters (Fig.~\ref{fig:exclusion}).

\section{Axions in Cosmology}
\label{sec:cosmology}

\subsection{Thermal axions and hot dark matter}

Axions are produced in the hot thermal plasma of the early universe
and, since they have a mass, contribute to the cosmic dark matter.
Before the QCD confinement transition at $T=\Lambda_{\rm QCD}$, the
relevant processes involve quarks and gluons, e.g., the gluonic
Primakoff effect caused by the axions's generic two-gluon
vertex~\cite{Turner:1986tb, Masso:2002np}. After confinement, the
most generic interaction process involves pions,
$\pi+\pi\leftrightarrow\pi+a$ \cite{Chang:1993gm}. Note that thermal
pions begin disappearing only at $T\alt30$~MeV, long after the QCD
epoch. Axions decouple after the QCD epoch if
$f_a\alt3\times10^7$~GeV, i.e., for $m_a\agt0.2$~eV.

Thermal axions provide a cosmic hot dark matter component fully
analogous to neutrinos. Therefore, the usual neutrino mass limits
from cosmic large-scale structure data can be adapted to the axion
case~\cite{Hannestad:2003ye, Hannestad:2005df}. Relying only on the
pion process for thermalization, for hadronic axions $m_a<1.05$~eV
(95\% CL), although this limit probably could be somewhat improved
with a more recent set of cosmological data. For comparison, we note
that this limit corresponds to $\sum m_\nu<0.65$~eV (95\% CL).

Axions or axion-like particles with a two-photon vertex decay into
two photons with a rate
\begin{eqnarray}
 \Gamma_{a\to\gamma\gamma}=\frac{g_{a\gamma}^2m_a^3}{64\pi}
 &=&\frac{\alpha^2}{256\,\pi^3}
 \left[\left(\frac{E}{N}-\frac{2}{3}\,\frac{4+z}{1+z}\right)
 \frac{1+z}{\sqrt{z}}\right]^2\,\frac{m_a^5}{m_\pi^2 f_\pi^2}
 \nonumber\\
 &=&1.1\times10^{-24}~{\rm s}^{-1}\,\left(\frac{m_a}{\rm eV}\right)^5
 \,,
\end{eqnarray}
where the first expression is general, the second applies to axions,
and the numerical one assumes $z=0.56$ and the hadronic case
$E/N=0$. Comparing with the age of the universe of
$4.3\times10^{17}$~s reveals that axions decay on a cosmic time
scale for $m_a\agt20$~eV. The decay photons would cause a variety of
observable consequences~\cite{Masso:1997ru}, depending on the axion
mass, so that cosmology alone entirely rules out axions with
$m_a>1$~eV (Fig.~\ref{fig:exclusion}). The low-mass end is covered
by the hot dark matter argument, seamlessly connecting to high
masses to avoid excess radiation from axion
decays~\cite{Masso:1997ru}.

Hot dark matter axions would be partially trapped in galaxies and
galaxy clusters. Searching for a decay line~\cite{Bershady:1990sw,
Ressell:1991zv} provides direct limits on a range of axion masses
marked ``Telescope'' in Fig.~\ref{fig:exclusion}.

\subsection{Non-thermal axions and cold dark matter}

The main cosmological interest in axions derives, of course, from
their possible role as the dominant cold dark matter component.
Almost immediately after ``invisible'' axion models had been
constructed it was recognized that they can be abundantly produced
by the ``misalignment mechanism'' \cite{Preskill:1982cy,
Abbott:1982af, Dine:1982ah}. After the spontaneous breaking of the
PQ symmetry at some high energy scale, the axion field relaxes to a
location somewhere in the Mexican hat potential. Near the QCD epoch,
instanton effects explicitly break the PQ symmetry, the very effect
that causes the dynamical PQ symmetry restoration. This ``tilting of
the Mexican hat'' causes the axion field to roll toward the
CP-conserving minimum, but this motion does not stop there. Rather,
coherent oscillations of the axion field are excited that ultimately
represent a condensate of cold dark matter. The cosmic matter
density in this homogeneous field mode is~\cite{Sikivie:2006ni}
\begin{equation}
 \Omega_ah^2\approx
 0.7\,\left(\frac{f_a}{10^{12}~{\rm GeV}}\right)^{7/6}\,
 \left(\frac{\Theta_{\rm i}}{\pi}\right)^2\,,
\end{equation}
where $-\pi\leq\Theta_{\rm i}\leq\pi$ is the initial ``misalignment
angle'' relative to the CP-conserving position. If the PQ symmetry
breaking takes place after inflation, $\Theta_{\rm i}$ will take on
different values in different patches of the universe. The
approximate average contribution is~\cite{Sikivie:2006ni,
Turner:1985si}
\begin{equation}
 \Omega_ah^2\approx
 0.3\,\left(\frac{f_a}{10^{12}~{\rm GeV}}\right)^{7/6}\,.
\end{equation}
Comparing with the cosmic cold dark matter density of $\Omega_{\rm
  CDM}h^2\approx 0.13$ this implies that axions with $m_a\approx
  10~\mu$eV would provide the dark matter (Fig.~\ref{fig:exclusion}).

However, this number sets only a crude scale of the expected mass
for axion dark matter. Apart from the overall particle-physics
uncertainties entering this result, the cosmological sequence of
events crucially matters. Assuming axions make up the cold dark
matter of the universe, significantly smaller masses are possible if
inflation took place after the PQ transition and the initial value
$\Theta_{\rm i}$ was relatively small. Conversely, if the PQ
transition takes place after inflation, there are additional sources
for nonthermal axion production, notably the formation and decay of
cosmic strings and domain walls associated with the breaking of the
PQ symmetry~\cite{Davis:1986xc, Sikivie:2006ni}. Therefore, the mass
of cosmic dark matter axions could be significantly larger than the
$10~\mu$eV scale. For a recent review of axion cosmology we refer to
Ref.~\cite{Sikivie:2006ni}.

If axions are the dark matter of our galaxy, they are accessible to
direct searches by the axion ``haloscope'' technique. A microwave
resonator is placed in a strong magnetic field and thus can be
driven by the galactic axion field due to the two-photon
coupling~\cite{Sikivie:1983ip, Bradley:2003kg}. After a series of
pilot experiments, the ADMX experiment in Livermore has reached the
axion line in a narrow range of masses~\cite{Asztalos:2003px}. The
project will be upgraded in several stages so that it should be able
to cover the range $1~\mu{\rm eV}\leq m_a\leq100~\mu$eV within the
upcoming decade. Therefore, axion dark matter can be detected in a
realistic range of parameters~\cite{Bibber:2006}.

In the exciting event that dark matter axions are found, intriguing
opportunities for observational axion cosmology open up. The fine
energy resolution that is possible with a microwave resonator allows
one to study the detailed axion velocity distribution that may
reveal narrow peaks from the phase-space folding that occurred
during galaxy formation~\cite{Duffy:2006aa, Sikivie:2006ni}.

\section{Summary}
\label{sec:summary}

Axions remain a well motivated solution of the strong CP problem.
Ongoing and near-future searches for solar axions by CAST and for
galactic cold dark matter axions by ADMX cover realistic parameters
and thus have a chance to detect these elusive particles. If CAST
were to detect solar axions, they would contribute a small hot dark
matter component of the universe so that these two experimental
directions are complementary. There remains a range of axion
parameters, roughly corresponding to $0.1~{\rm meV}\alt
m_a\alt10$~meV, where axions are not excluded, yet no detection
strategy has been proposed. Perhaps we will not have to worry about
this gap if axions show up in one of the accessible search ranges.

Unfortunately, the PVLAS signature likely has nothing to do with the
strong CP problem or dark matter. The particle interpretation is in
severe conflict with astrophysical arguments so that plausible
theoretical models are hard to construct. However, the effect is so
large that it can be tested relatively easily at several forthcoming
``shining light through walls'' experiments. Even if no new particle
is confirmed, a strong push is made towards measuring the
QED-implied birefringence of a magnetized vacuum. Such a measurement
would be an astounding achievement in itself.

\section*{Acknowlegments}

Partial support by the Deutsche Forschungsgemeinschaft under grant
SFB 375 and the European Union, contract RII3-CT-2004-506222
(ILIAS project), is acknowledged.

\section*{References}

\end{document}